# A comprehensive analysis of the (√13x√13)R13.9° type II structure of silicene on Ag(111)

**H Jamgotchian[1, \*], B Ealet[1], H Maradj[1, 2], J-Y Hoarau[1], J-P Bibérian[1] and B Aufray[1]**

[1] Aix-Marseille Université, CNRS, CINaM, UMR 7325, 13288 Marseille, France
[2] LSMC, Université d'Oran es-sénia, 31100, Oran, Algérie

\* E-mail: jamgotchian@cinam.univ-mrs.fr

**Abstract:** In this paper, using the same geometrical approach than for the (2√3x2√3) R30° structure (H. Jamgotchian et al., 2015, *Journal of Physics. Condensed Matter* **27** 395002), for the (√13x√13)R13.9° type II structure, we propose an atomic model of the silicene layer based on a periodic relaxation of the strain epitaxy. This relaxation creates periodic arrangements of perfect areas of (√13x√13)R13.9° type II structure surrounded by defect areas. A detailed analysis of the main published experimental results, obtained by Scanning Tunneling Microscopy and by Low Energy Electron Diffraction, shows a good agreement with the geometrical model.

**Keywords:** Silicene, Ag(111), superstructure, Moiré pattern, periodic deformation, strain relaxation.

**PACS numbers:** 62.23.Kn, 68.35.Gy, 68.37.Nq, 68.37.Ef

## Introduction

Silicene, the new allotrope of silicon equivalent to graphene, can be synthetized through silicon deposition on Ag [1–8] on Au(110) [9] and on Ir(111) [10] substrates, and by surface segregation on ZrB$_2$ thin film [11]. On Ag(111) different superstructures are observed due to a perfect match between four Ag-Ag interatomic distances and three second neighbors Si-Si interatomic distance. These superstructures correspond to different epitaxy relationships between the silicene layer with its honeycomb structure and the unreconstructed plane of the Ag(111) face. Experimentally, the superstructures strongly depend on the growth temperature and on the silicon deposition rate. They have all been characterized by LEED (Low Energy Electron Diffraction) and observed by STM (Scanning Tunneling Microscopy). It has been shown that the STM images of the different superstructures are related to the position of silicon atoms relative to the silver ones. In most cases, only the silicon atoms which are on top or close to the top of silver atoms are visible. Therefore, the positions of others Si atoms are subject to interpretation.

A geometrical approach shows that a simple rotation of the silicene layer relative to the silver substrate predicts all the observed superstructures [6]. Nevertheless, when the match between the silicene layer and the silver substrate is not perfect, the different superstructures are strained in compression or in expansion. This has been shown in different studies [12,13]. When the $(2\sqrt{3} \times 2\sqrt{3})R30°$ structure which is contracted, we showed in a previous paper that there is a periodic relaxation of the strain which gives rise to periodic defective areas located around perfect $(2\sqrt{3} \times 2\sqrt{3})R30°$ areas [14]. This generates Moiré-like patterns observed on STM images at low magnification and confirmed by LEED and FFT patterns through the apparition of extra spots instead of the main spots of the $(2\sqrt{3} \times 2\sqrt{3})R30°$ structure. This periodic relaxation also produces an average deviation of the superstructure angle (R30°) and a small decrease in the average periodicity.

In this paper, starting with an experimental STM image at high resolution, we develop an equivalent model for the $(\sqrt{13} \times \sqrt{13})R \pm 13.9°$ structure. In contrary to the $(2\sqrt{3} \times 2\sqrt{3})R30°$ structure, the $(\sqrt{13} \times \sqrt{13})R \pm 13.9°$ structure is strained in expansion relative to silver (see table 1 of [14]). A generalization of this model explains more accurately most of the STM images found in the literature: Moiré-like patterns and apparent global disorders.

**Atomic model of the perfect $(\sqrt{13} \times \sqrt{13})R13.9°$ type II structure.**

The $(\sqrt{13} \times \sqrt{13})R \pm 13.9°$ structures are of two kinds. The first one (type I) corresponds to $\alpha = \pm 27°$, and the second one (type II) to $\alpha = \pm 5.2°$ where $\alpha$ is the angle between the Si[110] and

Ag[110] directions [6,13]. In this paper, we focus our analysis on the (√13x√13)R±13.9° type II structure.

The geometrical model, generally accepted for the (√13x√13)R±13.9° type II structure produces four possible arrangements as shown on the figure 1. They differ by the α value and by the Si top atom environments. Figures 1(a) and 1(b) correspond to α = -5.2° (R = +13.9°) and α = + 5.2° (R = -13.9°) respectively [6,13,15]. In this work we limit ourselves to only in-plane models i.e. the buckling of the silicene layer is not taken into account. The relative position of Si atoms with respect to silver atoms are distinguished by the diameter and the brightness of the circles: e.g. Si atoms on top positions are the largest and brightest ones and correspond to the most visible atoms in STM images.

The four Si atoms, one on top and the three nearest neighbors, form a  shape (highlighted by a blue ), or form a  shape (highlighted by a green ). In figures 1(a) and 1(b), only  shapes are present, and in figures 1(c) and 1(d) only  shapes are present. Inside the unit cell (white diamonds), determined by Si top atoms, there are two nonequivalent areas: a dark one (highlighted by a dark grey circle) and a bright one (highlighted by a light grey circle). The dark area corresponds to four Si atoms close to three-fold Ag sites, and the bright area corresponds to six Si atoms, three close to top sites and three close to three-fold sites. The unit cells of figures. 1(a) and 1(c) (1(b) and 1(d) respectively) are not equivalent, they differ by the dark and bright areas positions inside the cell. On STM images,  and  shapes are not distinguishable. However, from the dark and bright areas positions inside the unit cell, we can determine the type of the silicon top atoms. It must be pointed out that the structures of figures 1(c) and 1(d) can be interpreted, with respect to the figures 1(a) and 1(b) respectively, as a translation of ¼ $d_{Ag}$ (0.0756nm) of silver unit cell, or as a rotation by 180° of the Si layer.

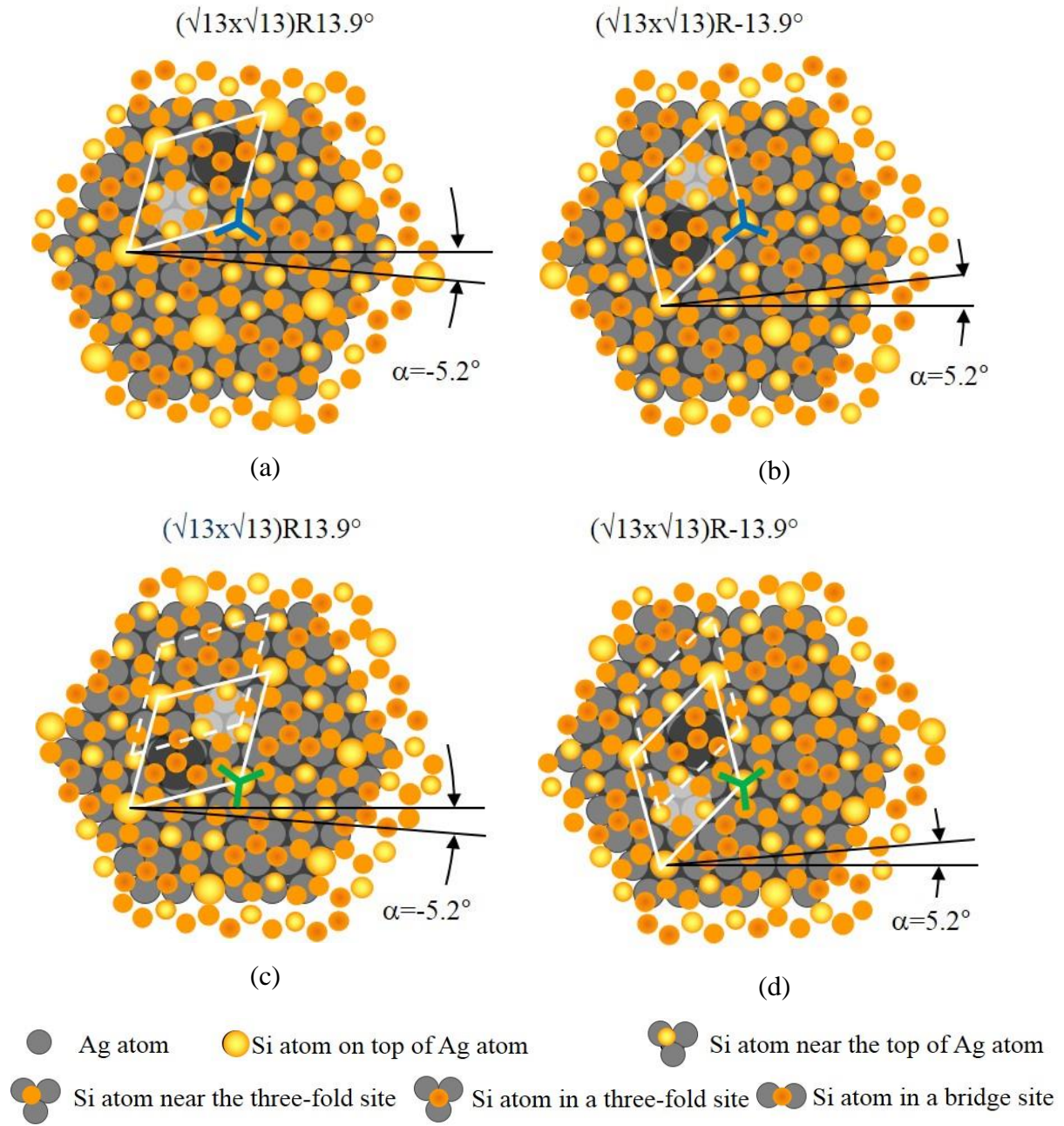

- ● Ag atom
- ● Si atom on top of Ag atom
- ● Si atom near the top of Ag atom
- ● Si atom near the three-fold site
- ● Si atom in a three-fold site
- ● Si atom in a bridge site

**Figure 1.** Atomic models of the (√13×√13)R±13.9° type II perfect structures. The white diamonds with solid lines correspond to the unit cells. (a) and (b) are (√13×√13)R±13.9° for α = ±5.2°. (c) and (d) are obtained by a translation of the silicene layer by ¼ $d_{Ag}$. The dashed white diamonds indicate the previous unit cell before translation. The same structure can be obtained by a rotation of the Si layer by 180°.

**Experimental observation of a local defect of a (√13x√13)R13.9°**

Experimentally, perfect (√13x√13)R13.9° structures are never observed in large view STM images. Figure 2(a) presents a high magnification STM image of the (√13x√13)R13.9° structure showing a defect area highlighted in green. The experimental protocol is given in a previous study [6]. Figure 2(b) shows the bare silver substrate from which we deduce the angle of the structure (-13.9°), and then, the angle $\alpha$ (−5.2°) of the silicene layer. In the high resolution STM image, the top part of the image corresponds to the perfect structure shown in figure 1(a). The bottom part corresponds to the perfect structure shown in figure 1(c). From this observation, we deduce that the defect area corresponds to the switch between green ⊥ and blue ⊥ shapes leading to a shift of the unit cells of ~0.2nm. We have observed this shift on many other STM images, which rules out the possibility of a thermal drift and/or piezo creep. Furthermore, such a linear defect has also been observed by M.R. Tchalala and al. [15], but interpreted as a 2D grain boundary. Here, we propose a model in which the defect area is due to a relaxation process.

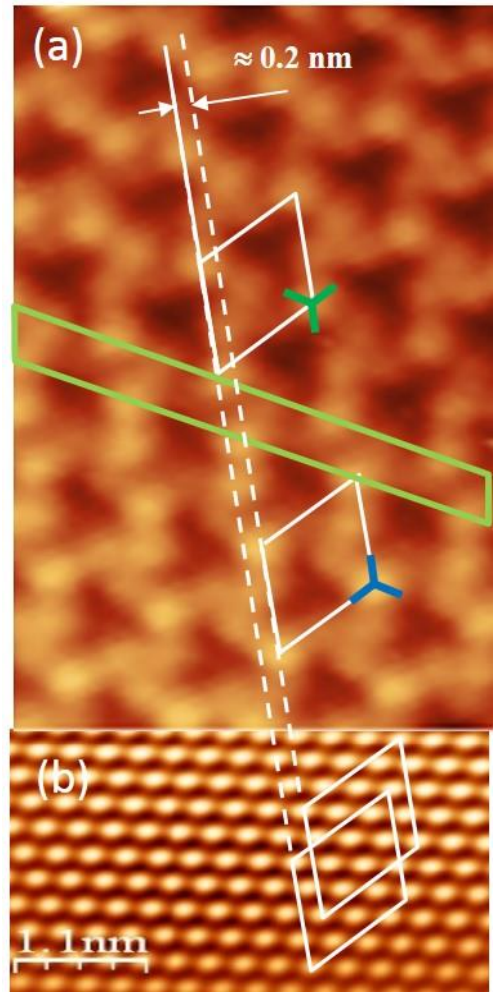

**Figure 2.** (a) High resolution STM image of the (√13x√13)R13.9° structure with a linear defect area highlighted by the green lines (I = 0.8 nA, V = -1.4 V, growth temperature: 270 °C and deposition rate 1.2 ML/h). (b) High resolution STM image of the bare silver substrate (I = 1.5 nA, V = -0.1 V)

## Atomic model of the (√13x√13)R13.9° type II structure with relaxation.

We have shown for the (2√3x2√3)R30° structure [14], that the Si layer strained in compression relaxes periodically by creating defective expanded areas. In contrast, the silicene layer having the (√13x√13)R13.9° structure being strained in expansion, should relax by creation of defective areas corresponding to a local contraction of the silicene layer.

As for the (2√3x2√3)R30° structure [14], we describe the mechanism in two steps: the linear defect formation, then the point defect formation. Figure 3 presents such a mechanism: the perfect ($\sqrt{13}\times\sqrt{13}$)R-13.9° structure is shown in figure 3(a) ($\alpha$ = +5.2° corresponding to figure 1(b)), along with the linear defect formation (figure 3(b)), and the point defect formation (figure 3(c)). The contraction can be obtained two ways: either by moving atom 1 along the blue arrow, or by moving atom 2 along the red arrow. We detail only the first possibility in the following model. The linear defect formation occurs in the area highlighted in green. The bottom part of figure 3(b) stays unchanged, and the top part is moved by approximatively ¼ $d_{Ag}$ in the direction of the blue arrows. In the defect area, for each Si atom, the mean displacement with respect to a perfect silicene layer (evidenced by dashed Si atoms) is only on average ~ 0.01 nm. The deformation changes a unit cell with Y shape to a unit cell with ⅄ shape. This corresponds to a change from the structure shown in figure 1(b) to the one of figure 1(d) with a shift of the unit cell. At the STM level, in the unit cell, the linear deformation switches the white areas into the black ones. On high resolution STM images, only the distance between dark areas can be measured and thus, near the defect area, only a shift of the black areas will be observed. The local increase in the distance between the two black areas induces an apparent very small increase of the average periodicity.

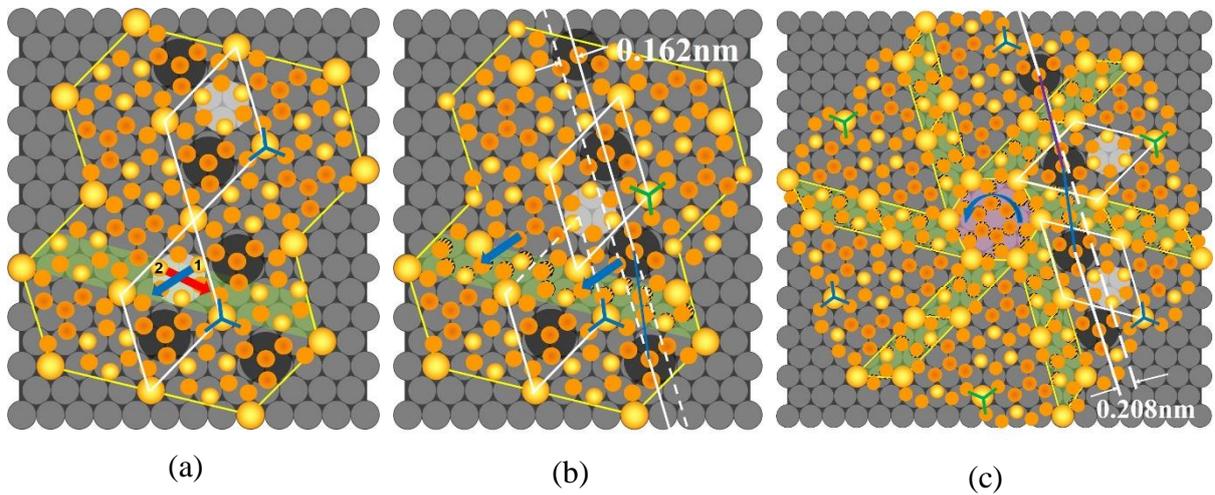

(a)　　　　　　　　　(b)　　　　　　　　　(c)

**Figure 3.** Schematic presentation of the formation of local defects in the ($\sqrt{13}\times\sqrt{13}$)R-13.9° structure for $\alpha$ = 5.2°. (a) Perfect ($\sqrt{13}\times\sqrt{13}$)R-13.9° structure. The green area indicates the place where the relaxation occurs. (b) In green, linear defects obtained by contraction of the silicene in the direction of the blue arrow ($\approx d_{Ag}/4$=0.07nm). The unit cell of the structure is symmetrically inverted and drifted after the contraction ($\sqrt{3}d_{Ag}$ = 0.50 nm) which induces an increase of the distance between two black areas on each side of the defect (0.162 nm in the direction perpendicular to the white lines). (c) In purple, point defects obtained by the combination of six linear defects (green) separating domains having alternatively ⊻ and ⋏ shapes. The Si atoms, visible by STM, form a hexagonal area (purple) which corresponds to a local (2√3x2√3)R30° structure. The shift between two identical black areas is 0.208 nm in the direction perpendicular to the white lines.

This deformation is described only for one direction, but it exists simultaneously in all the possible directions. The unit cell of the (√13x√13)R±13.9° structure has a three-fold symmetry and therefore, the relaxation is in three equivalent directions forming triangular areas. The intersection of these six triangular domains generates a point defect which has a six-fold symmetry (figure 3(c)) generating a local (2√3x2√3)R30° structure corresponding to a local positive rotation of the silicene layer (blue arrow). The point defect is surrounded by six perfect areas, corresponding alternatively to figure 1(b) and 1(d) (switch between ⊻ and ⋏ shapes). The shift between two nearest areas with the same unit cell is $d_{Ag}$ in the Ag[110] direction (figure 3(c)). According to our model, the direction of the rotation depends on the sign of α. In the case of α = -5.2°, the local rotation angle of the silicene layer will be negative.

We showed in a previous paper [14], that the unit cell of the (2√3x2√3)R30° structure has six-fold symmetry and therefore, the relaxation is in six equivalent directions producing hexagonal areas and a point defect with three-fold symmetry. In the case of the (√13x√13)R±13.9° structure, having three-fold symmetry, the relaxation is in three equivalent directions producing triangular areas and a point defect with six-fold symmetry.

A periodic extension of this model induces a superstructure shown in figure 4 (for α = +5.2°). There are three equivalent linear defects and three point defects, which surround the perfect areas. Equivalent linear defects are rotated by 120° with respect to the others.

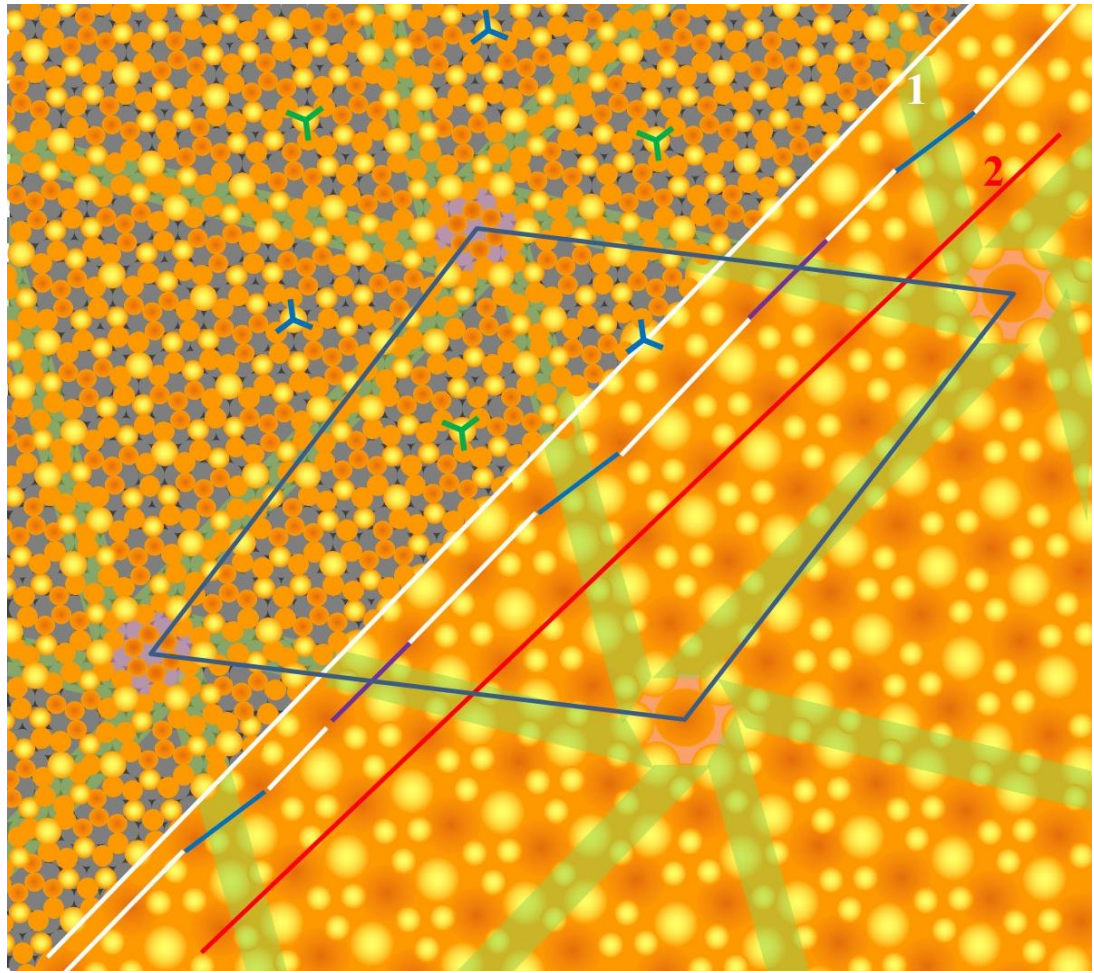

**Figure 4.** Large scale ball model of the (√13x√13)R-13.9° structure ($\alpha = +5.2°$) with periodic perfect areas surrounded by defects areas. The lower right part shows the expected STM image. The white-blue-purple zigzag line connects the different domains (perfect and defect areas). The white line corresponds to the orientation of a perfect (√13x√13)R-13.9° structure and the red line to the average angle which should be observed on STM images at large scale. The unit cell pattern (blue diamond) is a (√427x√427)R7.2° superstructure.

The upper left part presents the atomic ball model, and the lower right part represents the structure which should be observed with STM at low resolution. The tricolor white-blue-purple zigzag line connects the different domains (perfect and defect areas) as shown also in figure 3(c). The colors indicate different ways of crossing the defect areas. The shifts between two domains is 0.162 nm and 0.046 nm for blue and purple lines respectively, corresponding to a total shift of 0.208 nm, as indicated in figure 3(c). The white line corresponds to the orientation of a perfect (√13x√13)R13.9° structure and the red line to the average angle which would be determined on corresponding STM images.

Assuming that the average silicene hexagon distance remains at the bulk value (0.384 nm), the periodicity of the perfect and defective domains leads to a large (√427x√427)R7.2° superstructure (equivalent to a Moiré-like pattern) which is evidenced by the blue diamond in figure 4. The shift between the two white lines on the STM image of figure 2(a) (≈0.2 nm) is in good agreement with the distance predicted by the model (0.162 nm).

A dynamic analysis of the LEED patterns for such large superstructures being almost impossible to calculate, we used a simple kinematic approach, equivalent to a Patterson function [16,17], which is a convolution of the (√427x√427)R7.2° large superstructure unit cell with the perfect small local (√13x√13)R13.9° structure. As a consequence, on the LEED patterns, only the spots of the large superstructure close to the (√13x√13)R13.9° spots will be visible [18]. In fact both, the (√13x√13)R13.9° and the (√427x√427)R7.2° superstructures are commensurate with Ag(111), but not commensurate together.

Figure 5 is the expected LEED pattern for one domain of the (√427x√427)R7.2° superstructure ($\alpha$ = +5.2°). The yellow spots correspond to the silver substrate and the blue spots to the (√427x√427)R7.2° extra spots. The white line, which corresponds to the orientation of a perfect (√13x√13)R13.9° structure, crosses the doublets. The red line corresponds to the average angle measured on the STM image. It must be pointed out that the red line crosses some of the extra spots due to the superstructure. Experimentally, the LEED pattern will be a combination of the two domains ($\alpha$ = ±5.2°), as shown in the discussion section.

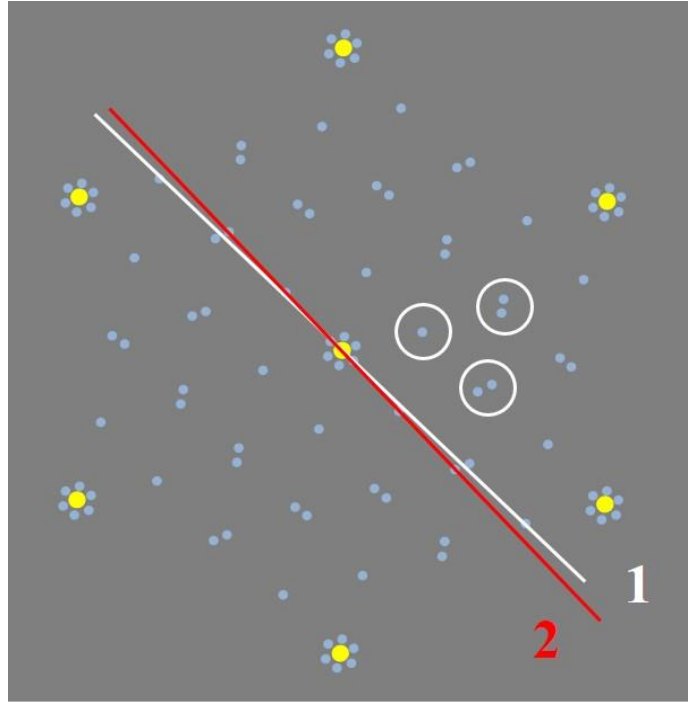

**Figure 5.** Expected LEED pattern for one domain of the (√427x√427)R7.2° superstructure (α = +5.2°) showing dots only near the dots of the (√13x√13)R13.9° structure.

**Generalization of the model**

This model shows that the periodic deviation creates an average rotation of the structure which value depends directly on the size of the perfect areas. By changing the size of the perfect (√13x√13)R13.9° areas, the size of the new superstructure and the angle shift will change: the larger the perfect domains, the smaller the angle shift and the structure will go towards a perfect (√13x√13)R13.9° structure. In other words, the smaller the number of defects, the smaller will be the angle shift. Experimentally, the angle shift can be measured on low-resolution STM images and the quantity of defects can be measured only on high-resolution STM images. The angle shift measurement allows estimating the average size of the perfect domains, which is equivalent to a statistical determination of the defects.

The model has been developed for the (√427x√427)R7.2° superstructure with triangular symmetry. The unit cell is constituted by two large triangles made of fifteen perfect (√13x√13)R13.9° unit cells. A simple superstructure with one (√13x√13)R13.9° unit cell would correspond to a (√63x√63)R16.6° superstructure. A generalization of the model allows us to predict the average angle (i.e. the angle shift) of the (√13x√13)R13.9° structure versus the distance between two perfect domains (i.e. superstructure periodicity) as it is shown in figure 6(a). The (√63x√63)R16.6° and (√427x√427)R7.2° superstructures are indicated by vertical dashed lines. On the calculated curves, we have reported the data issued from experimental studies.

Figure 6(b) shows the expected variations of the average size of the structure as a function of the superstructure periodicity. This curve shows that starting from the (√63x√63)R16.6° superstructure, which corresponds to the smallest superstructure periodicity, the average size of the structure should continuously decrease towards the perfect (√13x√13)R13.9° structure with infinite superstructure periodicity. For comparison, we have reported the experimental data issued from other studies (discussed in the following).

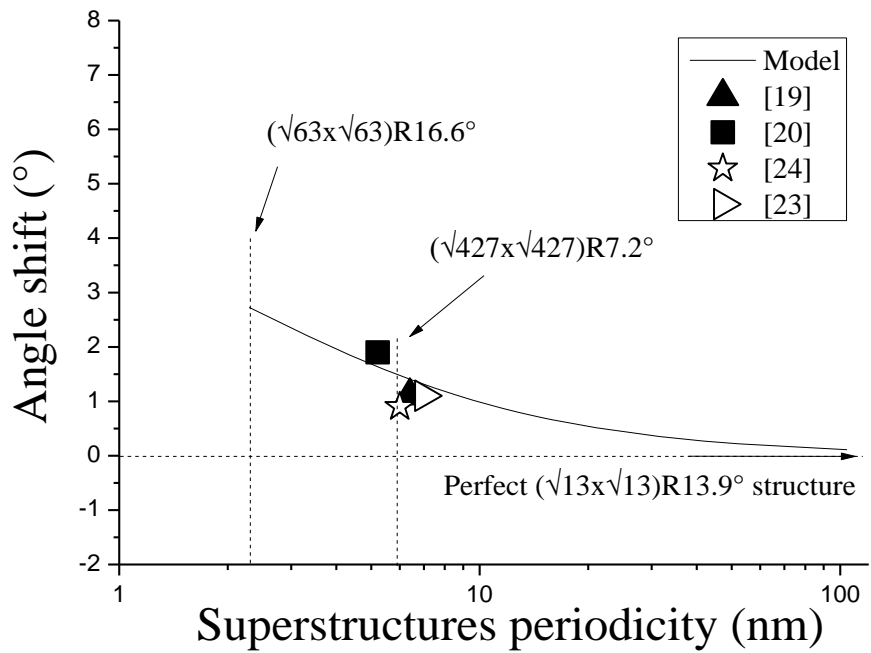

(a)

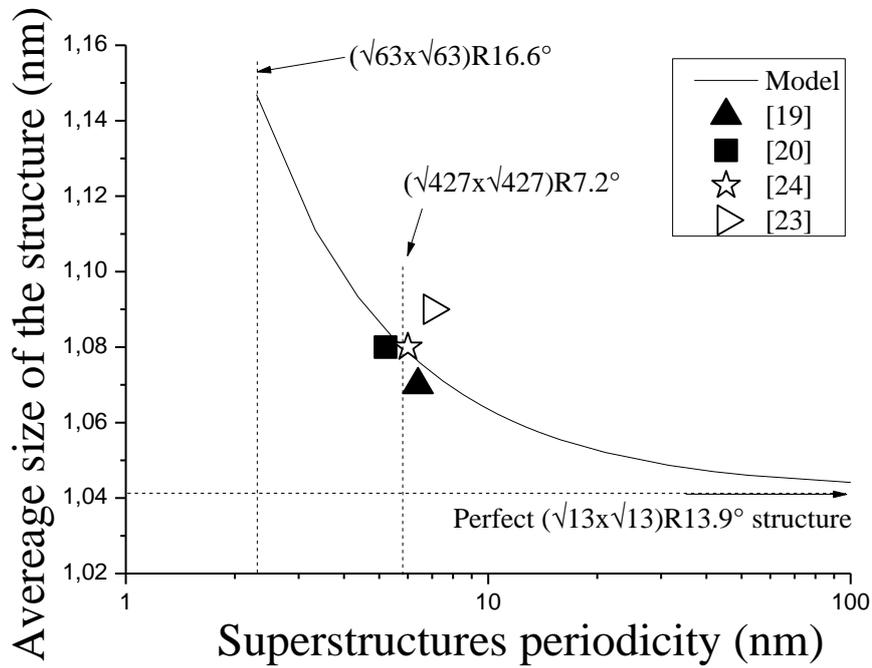

(b)

**Figure 6.** (a) Angle shift of the (√13x√13)R13.9° structure as a function of the superstructure periodicity. (b) Average size of the structure as a function of the superstructure periodicity for the (√13x√13)R13.9° structure.

## Discussion

Most studies assume that the (√13x√13)R13.9° structure is perfect over large domains [19,20] in agreement with DFT calculations [13,21,22]. Lin et al. [19] show a large view STM image that they consider as a perfect domain of (√13x√13)R13.9° structure, but a diffuse Moiré-like pattern can be identified. Recently, Liu et al. [20] have shown a very high quality STM image displaying a large area of (√13x√13)R13.9° structure. They also analyzed it as a perfect (√13x√13)R13.9° layer constituted of two interwoven domains which differ only by the z positions of the Si atoms, each domain being in agreement with the energetic DFT calculations of Guo et al. [22]. These domains produce periodic "vortexes". They all assume that the silicon honeycomb structure is perfect with respect to the Si hexagon distances without any defects. On the other hand, Tchalala et al. [15] have shown STM images with linear defects which have been interpreted by the formation of grain boundaries at the junction of two perfect

(√13x√13)R13.9° domains. As shown in the following, our geometrical model explains all the observed STM images of the (√13x√13)R13.9° type II structures [19,20,23,24].

We report in figure 7, the STM image of Lin et al. [19] showing two domains (4x4) and (√13x√13)R13.9° separated by a grain boundary. On the left side is a (4x4) structure which indicates the silver orientation. On the right side, the (√13x√13)R13.9° structure reveals a Moiré-like pattern which looks like wheels with six spokes limiting triangular domains. This is in excellent agreement with the model proposed in figure 4. We added in figure 7, a white-blue-purple zigzag line evidencing the local deviation of the structure in agreement with our model. The value for the average angle (14°) reported by the authors, corresponds to the measured angle in perfect area (blue arrow indicated by the authors). Averaging on the whole image (red line), we measured an average angle of ~13° and a superstructure periodicity of 6.4nm. These values reported on the curve of figure 6(a) are in good agreement with the model.

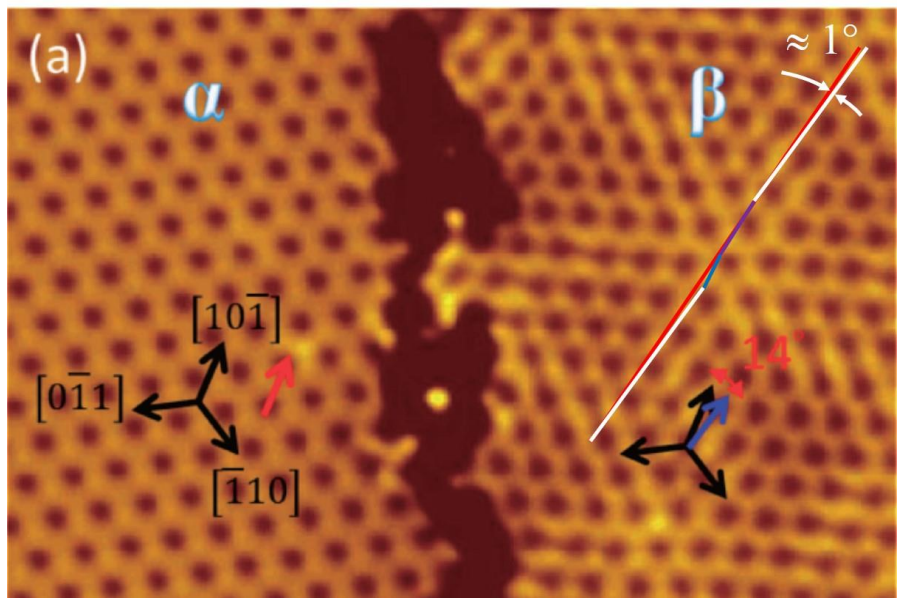

**Figure 7.** STM image of (4x4) and (√13x√13)R-13.9° structures given by Lin et al. [19]. We have measured on this image a superstructure periodicity of ~6.4nm, an average angle of ~13°, and an average periodicity of the structure of ~1.07nm.

We report in figure 8(a), the STM image of Liu et al. [20] showing a large area of (√13x√13)R13.9° structure constituted by wheels with six spokes delimiting triangular domains. This is again in excellent agreement with the model of figure 4, where the "vortexes" correspond to our point defects. We added in figure 8(a), a blue diamond evidencing the quasi-periodicity of the "vortexes". As in the model, the white line (#1) indicates the direction of the perfect ($\sqrt{13}$x$\sqrt{13}$)R13.9° structure and the red line (#2), the average direction of the observed structure. Figure 8(b) is a blow up of the white square, given by the authors. The red and black diamonds highlighted by the authors, evidence the two interwoven domains. We have highlighted by colors the areas corresponding to our model: in green the linear defects and in pink the point defect. The blue-white-purple zigzag line evidences the local deviation of the structure. In figure 8(c) we have calculated the FFT of figure 8(a). The white circles indicate the place where the ($\sqrt{13}$x$\sqrt{13}$)R13.9° spots should be. The color code for the lines is the same as in figure 8(a). The white line (#1, perpendicular to the one of figure 8(a)) crosses the middle of the doublets or triplets, and the red line (#2) passes on one of the spots of the doublets and triplets. The periodicity of the superstructure (~5.2 nm in agreement with the value of (5$\sqrt{13}$x5$\sqrt{13}$) given by the authors) and the average angle of the structure (~12°) are in good agreement with the FFT (figure 8(a)). These values reported on the curve of figure 6a are also in good agreement with the model. Note that the STM signature of the linear defect, highlighted in green, is composed of an alignment of squares (four spots with two brighter spots) exactly as predicted by the model in figure 4.

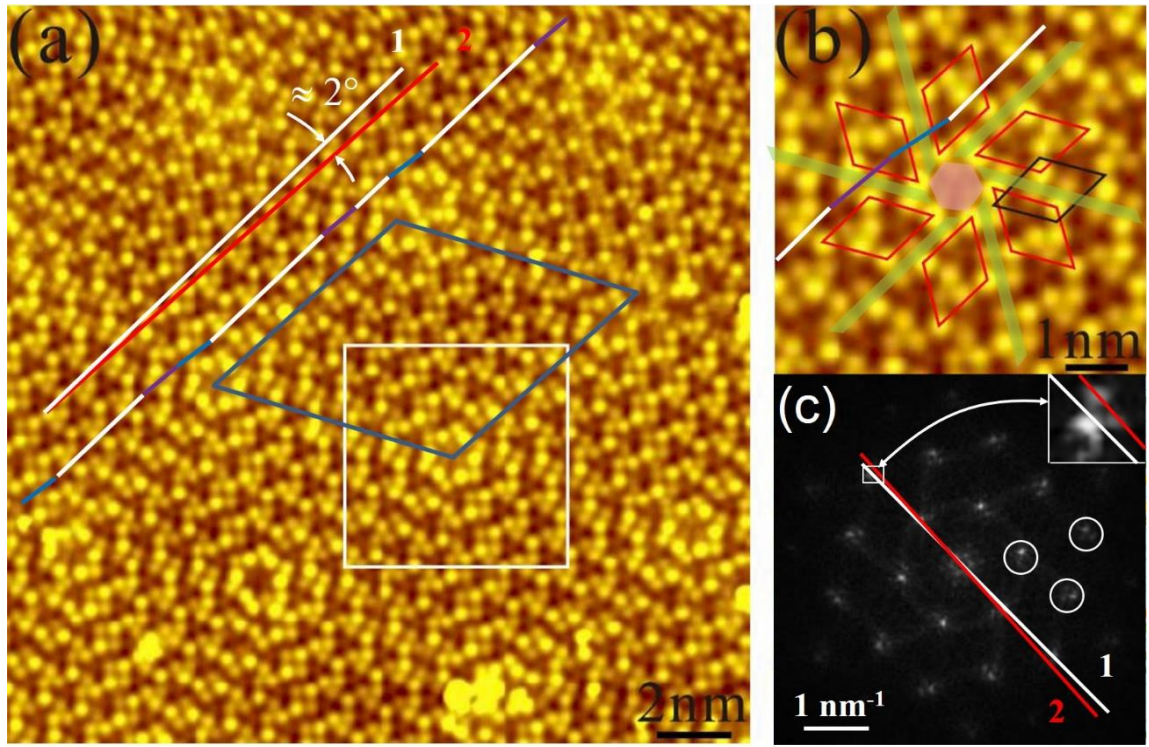

**Figure 8.** (a) (b) STM images of the (√13x√13)R13.9° structure from Liu et al. [20]. We have added white, red and tricolor zigzag lines, a blue diamond, green and purple areas corresponding to our model. (c) FFT of the STM image (a).

On figure 9(a), we present the LEED pattern corresponding to the STM image of figure 7 [19]. As a comparison, we present the LEED pattern (red dots) of a perfect (√13x√13)R13.9° structure in figure 9(b) and of the (√427x√427)R7.2° superstructure in figure 9(c), (both with the blue additional dots of the (4x4) structure and the yellow dots of the silver). On the experimental LEED, some weak extra dots, surrounded by white circles, are not observed on perfect structure. Those extra dots are generally not visible, but here the superstructure being well ordered these spots are present. On the LEED pattern, the positions of the extra spots are in agreement with the periodicity of the superstructure observed on the STM image (figure 7). On the FFT pattern shown in figure 8(c), doublets and triplets are also visible due to the quasi periodicity of the superstructure. Note that, the size of the STM image, 20x20nm$^2$, is comparable to the coherent width of the electron beam.

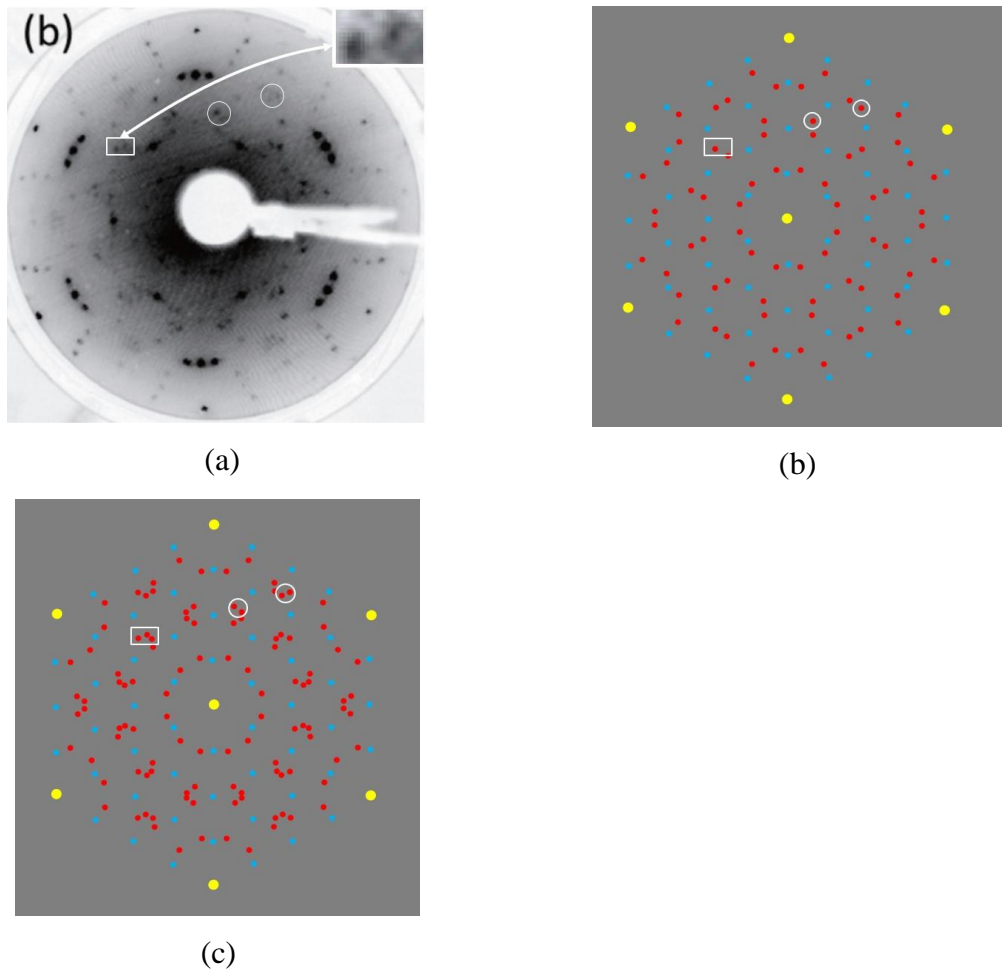

**Figure 9.** Experimental and calculated LEED patterns. (a) LEED pattern (53.9 eV) corresponding to the STM image of figure 7 of [19]. (b) Calculated LEED pattern of perfect (√13x√13)R13.9° (red dots) and (4x4) (blue dots) structures. (c) Calculated LEED pattern of the (√427x√427)R7.2° superstructure showing only dots near perfect (√13x√13)R13.9° structure. The yellow dots correspond to silver substrate.

The geometrical model is in line with all the experimental observations:
- The STM images of large domains of (√13x√13)R13.9° structure show Moiré-like patterns which look like wheels with six spokes limiting triangular areas.
- The STM signatures of the spokes are formed by alignment of squares constituted by four spots with two brighter spots.

The LEED pattern and the FFT obtained from the STM image exhibit extra dots due to the superstructure.

Another fact supporting the model is that the same methodology (periodic areas strained and relaxed), allow the explanation of the STM images and LEED patterns for the (2√3x2√3)R30° and (√13x√13)R13.9° structures. The difference between the two structures takes place in the nature of the strain and in the symmetry of their unit cells. In the perfect areas, the (2√3x2√3)R30° is strained in compression whereas the (√13x√13)R13.9° is strained in expansion. It is the opposite in the defect areas. On the STM images, the hexagonal symmetry of the unit cell of the (2√3x2√3)R30° structure leads to perfect domains with hexagonal shapes, whereas for the (√13x√13)R13.9°, the symmetry of third order of the unit cell induces the formation of perfect domains with triangular shapes.

Interestingly, the point defects of the (√13x√13)R13.9° structure appear as a local area of (2√3x2√3)R30° structure and the point defects of the (2√3x2√3)R30° structure appear as local area of (√13x√13)R13.9° structure. For the two structures ((2√3x2√3)R30° and (√13x√13)R13.9°), the variations in the average periodicities, due to the defects, are too small to put in evidence the periodic relaxation phenomena. However, the deviation angle observed experimentally is between 4 and 8° for the (2√3x2√3)R30° structure [14] and only between 1 and 2° for the (√13x√13)R13.9° structure. This could be the reason why the (√13x√13)R13.9° structure is often assumed to be perfect unlike the (2√3x2√3)R30° structure which is often denied [25].

**Conclusion**

In this paper, as for the (2√3x2√3)R30° structure, we show, similarly, that the (√13x√13)R13.9° structure is not perfect but composed of a periodic arrangement of perfect and defective areas. In the model, the perfect areas of silicene layers are slightly expanded due to a strain epitaxy, whereas the defective areas are attributed to a local relaxation of this strain. This model explains perfectly well the STM images of large areas showing Moiré-like pattern which looks like wheels with six spokes limiting triangular domains. The spokes are composed of an alignment of squares (four spots with two brighter spots) exactly as predicted by the model. As for the (2√3x2√3)R30° structure, this shows that despite the impression of disorder given by the STM images, the silicene film remains a continuous honeycomb layer with only local and small periodic deformations. The proposed model is geometrical and not theoretical.

As for the (2√3x2√3)R30° structure the layer is continuous with topological defects. This explains why the field effect transistor works as well for a (2√3x2√3)R30° structure than for a mix of (4x4) and (√13x√13)R13.9° structures [26].